\documentclass[11pt]{article}

\usepackage[utf8]{inputenc}
\usepackage{balance}

\usepackage{amsmath, amssymb, graphicx}
\usepackage{geometry}
\usepackage{graphicx}
\usepackage{caption}
\usepackage{amsthm}
\usepackage{float}
\usepackage{appendix}
\usepackage{verbatim}
\usepackage{epstopdf}
\usepackage{amssymb}
\usepackage{subfigure}
\usepackage{subcaption}
\usepackage{color}
\usepackage{cite}
\usepackage{cancel}
\usepackage{amsthm,mathrsfs,amsfonts,dsfont}
\DeclareMathOperator{\sinc}{sinc}

\newtheorem{remark}{Remark}
\newtheorem{corollary}{Corollary}

\newtheorem{lemma}{Lemma}

\newtheorem{definition}{Definition}
\usepackage[shortlabels]{enumitem}

\usepackage{listings}
\usepackage{color}
\definecolor{mv}{rgb}{1,0.5,0} 
\definecolor{codegreen}{rgb}{0,0.6,0}
\definecolor{codegray}{rgb}{0.5,0.5,0.5}
\definecolor{codepurple}{rgb}{0.58,0,0.82}
\definecolor{backcolour}{rgb}{0.95,0.95,0.92}

\usepackage{xcolor} 
\begin{document}

\title{Privacy via Modulation Rotation and Inter-Symbol Interference}

\author{
    Morteza Varasteh\thanks{School of Computer Science and Electronic Engineering, University of Essex, Colchester, UK. Email: m.varasteh@essex.ac.uk} \and
    Pegah Sharifi\thanks{Amirkabir University of Technology, Tehran, Iran. Email: p.sharifi96@aut.ac.ir}
}


\maketitle

\begin{abstract}
Two physical-layer mechanisms for achieving user-side differential privacy in communication systems are proposed. Focusing on binary phase-shift keying (BPSK) modulation, differential privacy (DP) is first studied under a deterministic phase rotation applied on the BPSK modulation at the transmitter, while the receiver is assumed to be unaware of the rotation angle. In this setting, privacy is achieved through an effective reduction in the decision distance, resulting in a controlled increase in the bit error rate (BER) without explicit noise injection. Next, a BPSK transmission scheme with intentionally induced inter-symbol interference (ISI) is studied, where the receiver is likewise unaware of the deterministic timing offset that generates the ISI. Unlike the rotated BPSK scheme, the DP obtained via ISI is shown to depend explicitly on the input data distribution. In particular, numerical results demonstrate that, for a fixed ISI parameter, the privacy loss is maximized when the binary input symbols are equiprobable. While conventional DP mechanisms rely on artificially added noise—often incurring additional energy or communication costs, it is shown that structured modifications, such as modulation rotation or induced ISI inherent to realistic communication channels can itself provide DP guarantees. While the analysis focuses on deterministic transmitter modifications unknown to the receiver, it is noted that real-world devices naturally introduce unintentional rotations or ISI due to hardware nonidealities and implementation errors. These effects can therefore provide a level of privacy without requiring explicit noise injection. Hence, it is possible to avoid deliberately perturbing the data, instead leveraging inherent device imperfections to achieve privacy guarantees with no additional privacy cost.
\end{abstract}

\section{Introduction}

Ensuring data privacy in communications and learning systems is a growing challenge in modern network infrastructures. 
Applications such as environmental monitoring, Internet of Things (IoT), and federated learning (FL) rely on numerous devices that share locally observed information through wireless links. 
While these systems enable large-scale intelligence, they also risk exposing sensitive user data. 
Differential privacy (DP) provides a formal mathematical guarantee against such disclosure by ensuring that small changes in the underlying data produce statistically indistinguishable outputs. 
However, conventional DP mechanisms depend on the addition of artificial noise (either at the device or server side), which consumes energy \cite{DPnoise1}. 

Recent research has shown that the inherent stochasticity of wireless channels, such as thermal noise, fading, and superposition interference, can itself serve as a privacy-preserving mechanism.  
Liu and Simeone~\cite{liu2020privacy} first demonstrated that in wireless federated learning with uncoded analog aggregation, the channel noise alone can guarantee a specified DP level below a certain signal-to-noise (SNR) threshold, coining the notion of \emph{privacy for free}.  
Building on this principle, Li \emph{et al.}~\cite{li2025free} proposed a channel-native bit-flipping strategy that leverages communication errors as a DP process, while Liu \emph{et al.}~\cite{liu2023mimo} extended the analysis to MIMO fading channels and examined the role of spatial diversity.  
Liang \emph{et al.}~\cite{liang2025perk} formalized the view of ``differential privacy as a perk,'' showing that multi-access fading channels with optimized beamforming can inherently satisfy DP without explicit randomization.  
Yan \emph{et al.}~\cite{yan2023device} further incorporated receiver noise into device scheduling policies for over-the-air FL, demonstrating how channel noise can be managed as a controllable privacy resource.

Complementary studies have analyzed related mechanisms where natural receiver noise or interference replaces artificial perturbation.  
Results on AirComp-based FL have characterized DP obtained directly from channel superposition~\cite{liu2023mimo,liang2025perk}, while others have explored power-adaptive transmission schemes that harness receiver noise for privacy preservation without sacrificing learning performance.  
Together, these studies establish the principle that privacy may emerge as a byproduct of physical-layer randomness rather than a software-enforced constraint.
 
The ``privacy for free'' idea resonates with foundational concepts in physical-layer security, where intrinsic channel randomness has long been viewed as a source of secrecy.  
Rabbachin \emph{et al.}~\cite{rabbachin2015intrinsic} formalized \emph{intrinsic network secrecy}, showing that aggregate interference and spatial randomness can ensure confidentiality without key exchange.  
Wei \emph{et al.}~\cite{wei2021anon} proposed \emph{anonymous precoding}, where randomized precoder design conceals the sender’s identity to achieve physical-layer anonymity.  
Similarly, Zoli \emph{et al.}~\cite{zoli2020pls} demonstrated that time–frequency channel entropy can be transformed into shared keys without cryptographic infrastructure.  
These works highlight a unifying principle: privacy and security can be obtained directly from the natural uncertainty of the wireless medium.

Despite these advances, most ``free privacy'' studies consider only additive white Gaussian noise (AWGN), fading, or aggregate interference as sources of randomness.
In practice, however, communication links exhibit additional ubiquitous and structured uncertainties originating not only from propagation but also from transmitter hardware nonidealities. 
One prominent example is constellation (modulation) rotation, which can arise unintentionally from carrier-frequency offsets, oscillator phase noise, residual phase errors, or I/Q imbalance in the RF front-end, yielding an effective unknown phase rotation of the transmitted symbols at the receiver. Such rotation is typically treated as an impairment to be estimated and compensated, but when it is unknown, it contributes an additional layer of obfuscation.
Communication channels also exhibit another ubiquitous and structured form of uncertainty: \emph{inter-symbol interference} (ISI). ISI arises from pulse shaping, synchronization offsets, and multipath propagation, and is traditionally regarded as a distortion to be mitigated. Yet, from a privacy standpoint, ISI acts as a signal-dependent perturbation that inherently obscures the transmitted data. To date, no study has rigorously analyzed how the joint effect of unintentional transmitter-induced modulation rotation and temporal ISI can be quantified and harnessed as a natural source of DP, potentially enabling privacy gains without deliberately injecting additional noise.

In this paper, we introduce a new perspective by showing that ISI, together with constellation rotation, can be characterized and exploited to achieve DP at the user side.
Rather than treating ISI and phase rotation solely as impairments to be eliminated, we model them as structured and signal-dependent perturbations that act as intrinsic privacy resources when they are unknown at the receiver.
By establishing analytical relationships between ISI strength, phase-rotation uncertainty, signal-to-noise ratio, and the DP parameter~$\varepsilon$, we demonstrate that meaningful privacy guarantees can be obtained without injecting artificial noise.

Notations: The \textbf{Hurwitz zeta function} \(\zeta(s, q)\) is defined for complex \(s\) with \(\Re(s) > 1\) and real \(q > 0\) by the series
\[
\zeta(s, q) = \sum_{n=0}^{\infty} \frac{1}{(n + q)^s}.
\]
The \textbf{alternating Hurwitz zeta function}, also known as the Hurwitz-type Euler zeta function, is given by
\[
\zeta^*(s, q) = \sum_{n=0}^{\infty} \frac{(-1)^n}{(n + q)^s},
\]
where \(s\) is a complex number with \(\Re(s) > 1\) and \(q\) is a real number.

\section{System Model}
We consider a point-to-point communication system impaired by complex AWGN. 
The transmitter emits a baseband signal $\pmb{x}(t)$ given by
\begin{equation}\label{TxSignal}
\pmb{x}(t) = \sum_{n} \pmb{a}_{n} e^{j\alpha}\, q(t - nT - \tau),
\end{equation}
where $\pmb{a}_{n} \in \{+1, -1\}$ denotes the $n$th binary symbol with $\Pr\{\pmb{a}_{n} = +1\} = p$. 
The parameter $\tau$ and $\alpha$ represent a deterministic timing offset and the phase shift, respectively, introduced during transmission and is assumed to be unknown to the receiver\footnote{We express the transmitted signal in~(\ref{TxSignal}) in a generic form that encompasses both parameters $\tau$ and $\alpha$. In the subsequent analysis, we study the effect of each parameter separately by setting the other parameter to zero.}. 
The function $q(t)$ denotes the unit-energy pulse shape, which may be a sinc, raised-cosine, or any other band-limited waveform, and $T$ corresponds to the Nyquist symbol interval.

After passing the received signal through a filter matched to $q(\cdot)$, the output is given by
\begin{equation}
\pmb{y}(t) = \sum_{n} \pmb{a}_{n}e^{j\alpha}\, g(t - nT - \tau) + \pmb{n}(t),
\end{equation}
where $g(\cdot)$ denotes the impulse response of the matched filter corresponding to the transmit pulse $q(\cdot)$, and $\pmb{n}(t)$ represents the filtered complex Gaussian noise component. 
The receiver samples $\pmb{y}(t)$ at intervals of $T$ seconds, yielding
\begin{align}\label{eq:0}
\pmb{y}_{m} &= \pmb{y}(mT) = \sum_{n} \pmb{a}_{n}e^{j\alpha}\, g((m - n)T - \tau) + \pmb{n}(mT) \\
\label{eq:5}
&= \pmb{a}_{m}e^{j\alpha}\, g(-\tau) 
   + \sum_{n \neq m} \pmb{a}_{n}e^{j\alpha}\, g((m - n)T - \tau)
   + \pmb{n}(mT),
\end{align}
where the first term represents the desired symbol component, while the second term captures the ISI introduced by non-zero offset~$\tau$.

Throughout this work, we analyze~\eqref{eq:5} under two complementary scenarios: a fixed, nonzero phase rotation $(\tau = 0, \alpha \neq 0)$ and a fixed, nonzero timing offset $(\tau \neq 0, \alpha = 0)$. In each case, the nonzero parameter is assumed to remain constant over the entire data transmission.
For notational simplicity, the time index $m$ is omitted in what follows, yielding
\begin{align}\label{eq:6}
\pmb{y} &= \pmb{a}e^{j\alpha}\, g(\tau) + \sum_{t \neq 0} \pmb{a}_{t}e^{j\alpha}\, g(tT - \tau) + \pmb{n} \\
&= \pmb{a}e^{j\alpha}\, g(\tau) + \mathbf{I} + \pmb{n},
\end{align}
where $g(\cdot)$ denotes the pulse-shaping function, typically chosen as a sinc or raised-cosine waveform, for which $g(-\tau) = g(\tau)$. 
Here, $\pmb{a}$ represents the BPSK-modulated symbol, $\mathbf{I}$ captures the ISI component in~\eqref{eq:6}, and $\pmb{n}$ is the complex Gaussian noise term with zero mean and variance~$\sigma^{2}$, i.e., $\pmb{n} \sim \mathcal{CN}(0, \sigma^{2})$. 
Unless otherwise stated, the symbol duration is normalized to $T = 1$ throughout this work.

In the following, we study DP at the bit level from the receiver’s perspective. 
For completeness, the definition of differential privacy is recalled below.

\begin{definition}[Differential Privacy]
Let $\mathcal{D}_1$ and $\mathcal{D}_2$ be two \emph{adjacent datasets} that differ in only one entry. 
A randomized mechanism $\mathcal{A}$ is said to satisfy $\varepsilon$-\emph{differential privacy} if, for all measurable subsets $S$ of the output space,
\begin{equation}
\Pr\{\mathcal{A}(\mathcal{D}_1) \in S\}
    \le e^{\varepsilon}
       \Pr\{\mathcal{A}(\mathcal{D}_2) \in S\},
\end{equation}
where $\varepsilon \ge 0$ is the \emph{privacy budget}. 
A smaller value of $\varepsilon$ indicates a stronger level of privacy, as the outputs of $\mathcal{A}$ become increasingly indistinguishable when generated from neighboring datasets.
\end{definition}

In the context of this work, the mechanism $\mathcal{A}$ corresponds to the stochastic mapping induced by the communication channel and receiver operation on the transmitted bit sequence. 
Hence, the effective privacy arises from the combined randomness of channel noise and either of phase rotation or ISI, rather than from any artificially injected perturbation.

\section{Main results}
In the following, we first briefly review one simple scenario, namely, \emph{privacy via SNR}.
Afterwards, we focus on \emph{privacy via a deterministic phase rotation} and \emph{privacy via a deterministic offset}.

\subsection{Privacy via SNR}
Assume $\alpha =\tau = 0$ in~\eqref{eq:6}. 
In this case, the received signal reduces to $\pmb{y} = \pmb{a} + \pmb{n}$. 
For any given privacy budget~$\varepsilon$, the optimal\footnote{Optimality is in the min--max sense for the probability of error, i.e., the minimum of the maximum probability of error, where the maximum is taken over all possible input distributions. 
This worst-case formulation is standard in DP analysis, where the mechanism is assumed to be agnostic to the input distribution.} 
binary-input binary-output mechanism that satisfies the privacy loss~$\varepsilon$ is a binary symmetric channel (BSC) with crossover probability
\begin{equation}\label{eq:bsc_opt}
q^\star(\varepsilon) = \frac{1}{1 + e^{\varepsilon}}.
\end{equation}

When a message is transmitted via BPSK modulation, the bit error rate (BER) over an AWGN channel is given by
\begin{equation}\label{eq:ber}
p_e = Q\!\left(\frac{\sqrt{E_b}}{\sigma}\right),
\end{equation}
where $E_b$ denotes the bit energy, and $Q(\cdot)$ is the Gaussian $Q$-function.\footnote{$Q(x) = \frac{1}{\sqrt{2\pi}} \int_x^{\infty} e^{-t^2/2}\,dt$. 
Throughout this work, and without loss of generality, we assume $E_b = 1$.}

Hence, the mapping from the input bit to the decoder output is equivalent to a BSC with crossover probability $p_e$, namely BSC($p_e$), resulting in an effective privacy budget\footnote{It is assumed that the receiver is \emph{honest-but-curious} and adheres to the prescribed decoding protocol. 
Otherwise, a malicious receiver could intentionally misreport or manipulate the effective SNR, 
in which case privacy based solely on channel noise could be trivially compromised.}

\begin{equation}\label{eq:eps_snr}
\varepsilon = \ln\!\left(\frac{1 - p_e}{p_e}\right),
\qquad
p_e = Q\!\left(\frac{1}{\sigma}\right).
\end{equation}

If the resulting $\varepsilon$ in~\eqref{eq:eps_snr} satisfies the target privacy level, no additional processing is required. 
Otherwise, a conventional approach is to introduce an additional layer of artificial noise before transmission.

\textbf{Conventional approach:} 
Pass the binary message through a BSC($\alpha$) prior to BPSK modulation. 
The end-to-end channel then becomes a cascade of BSC($\alpha$) and BSC($p_e$), equivalent to a single BSC($\alpha * p_e$), where
\begin{equation}
\alpha * p_e \;\triangleq\; \alpha(1 - p_e) + p_e(1 - \alpha)
\end{equation}
denotes the binary convolution of mass probabilities. 
By tuning $\alpha$, the desired $\varepsilon$ can be achieved according to
\begin{equation}
\varepsilon = \ln\!\left(\frac{1 - (\alpha * p_e)}{\alpha * p_e}\right).
\end{equation}

This method, referred to as \emph{privacy via SNR}, avoids explicit noise addition in the signal domain and instead adjusts the energy per bit to modulate the bit error rate, thereby influencing the achievable privacy level~$\varepsilon$.

\subsection{Privacy via a Deterministic Phase Rotation}
Consider a single-user transmission with a fixed phase rotation~$\alpha$ applied to the transmitted BPSK symbols. 
The received signal is given by
\begin{equation}
\pmb{y}_{\alpha} = e^{j\alpha}\pmb{a} + \pmb{n}.
\end{equation}
The receiver is assumed to perform coherent detection along the real axis, while being unaware of the phase rotation~$\alpha$.

Under this condition, the projection of the received symbol onto the detection axis reduces the effective decision distance by a factor of $\cos\alpha$. 
Accordingly, the bit error probability is expressed as
\begin{equation}\label{eq:pe_phase}
p_e(\alpha) = 
Q\!\left(\frac{\,\cos\alpha}{\sigma}\right).
\end{equation}

This mechanism achieves privacy without adding explicit noise; instead, it modulates the phase of the transmitted symbol, thereby reducing the effective SNR observed at the receiver. 
For a given~$\alpha$, the corresponding privacy level can be obtained by substituting $p_e(\alpha)$ from~\eqref{eq:pe_phase} into
\begin{equation}\label{eq:eps_phase}
\varepsilon(\alpha) = \ln\!\left(\frac{1 - p_e(\alpha)}{p_e(\alpha)}\right).
\end{equation}
Conversely, to achieve a target privacy budget~$\varepsilon$, the required phase rotation can be determined by inverting~\eqref{eq:pe_phase}:
\begin{equation}\label{eq:alpha_required}
\cos\alpha = \frac{\sigma}{\sqrt{E_b}}\,Q^{-1}\!\left(\frac{1}{1 + e^{\varepsilon}}\right).
\end{equation}

If the obtained $\varepsilon$ from~\eqref{eq:eps_phase} meets the desired privacy requirement, no further adjustment is needed. 
Otherwise, the user can increase~$\alpha$ to enhance privacy. 
Note that for $\alpha = 0$, the signal is unrotated, corresponding to maximum SNR and maximum privacy loss; whereas for sufficiently large $\alpha$ (approaching $\pi/2$), the received energy along the detection axis vanishes, leading to complete indistinguishability and $\varepsilon \approx 0$, namely perfect privacy.

\begin{remark}
Note that both privacy via phase rotation and privacy via SNR is independent of the input distribution $p$. Also note that privacy via phase rotation simplifies to privacy via SNR when $\alpha=0$. 
\end{remark}

\subsection{Privacy via a Deterministic Offset}

We now consider the case of a single user with zero phase rotation but a nonzero timing offset~$\tau$. 
The received signal can be expressed as
\begin{align}
\pmb{y}_{\tau} &= \pmb{a}\, g(\tau) + \mathbf{I} + \pmb{n}, 
\end{align}
where the first term corresponds to the desired symbol component, and $\mathbf{I}$ represents the ISI resulting from the overlap of neighboring symbols due to the offset~$\tau$. 
Explicitly, $\mathbf{I}$ is written as
\[
\mathbf{I} = \sum_{t \neq 0} \pmb{a}_{t}\, g(tT - \tau).
\]
The ISI term~$\mathbf{I}$ can be interpreted as a weighted sum of random BPSK symbols, where the weights are determined by the pulse-shaping function~$g(\cdot)$ and the offset~$\tau$. 
This naturally introduces a random perturbation at the receiver side, even in the absence of additive channel noise.

From Kolmogorov’s three-series theorem~\cite[Theorem~22.8]{Bill95}, 
the random variable~$\mathbf{I}$ (representing the ISI contribution for one received sample) converges in~$\mathds{R}$ with probability one, since 
$-1 \le \pmb{a}_t\, g(t - \tau) \le 1$ for all~$t$ almost surely, 
and both $\sum_{t \neq 0} g(t - \tau)$ and $\sum_{t \neq 0} g(t - \tau)^2$ converge. 
Hence, the ISI process is well-defined and admits finite moments.  
The following lemma characterizes the sum of pulse coefficients with different powers for a commonly used pulse shape.

\begin{lemma}\label{lemma1}
    When sinc function is used as the pulse, i.e., $g(\tau) = \sinc(\tau) =\frac{\sin{\pi\tau}}{\pi \tau}$, and $\tau\in(0,1)$ we have
    \begin{align}\label{eq:lemma}
    G(n, \tau) \triangleq\sum_{t\neq 0}g(t-\tau)^n &= \begin{cases} 
  1-\sinc(\tau) & \text{if } n =1 \\
  \sinc(\tau)^n(\tau^n(\zeta(n, \tau)+\zeta(n, -\tau))-2) & \text{if }  n \text{ is even}\\
  \sinc(\tau)^n(\tau^n(\zeta^*(n, \tau)-\zeta^*(n, -\tau))-2) & \text{if } n \text{ is odd}
\end{cases}
\end{align}
\end{lemma}
\textit{Proof}: See Appendix \ref{App1}

The result in Lemma~\ref{lemma1} ensures that the aggregated ISI component~$\mathbf{I}$ admits a well-behaved distribution whose moments depend smoothly on the offset~$\tau$. 
We can thus compute the central and raw moments of~$\mathbf{I}$, which are key to quantifying its contribution to DP.

\begin{corollary}\label{cor1}
All central moments of the random variable $\mathbf{I}$, i.e.,
\[
\mu_n \triangleq \mathds{E}\!\left[(\mathbf{I} - \mathds{E}[\mathbf{I}])^n\right], \quad n \ge 0,
\]
are finite and given by
\begin{align}
\mu_n = G(n, \tau)\, 2^n \!\left[p(1 - p)^n + (1 - p)(-p)^n\right].
\end{align}
Let $\mu'_n \triangleq \mathds{E}[\mathbf{I}^n]$ denote the raw moments, and define $\mu \triangleq \mu'_1 = (2p - 1)(1 - \sinc(\tau))$. 
Then, for $n \ge 2$, the higher-order raw moments are obtained as
\begin{align}
\mu'_n = \sum_{j = 0}^{n} \binom{n}{j} \mu_j \, \mu^{n - j}.
\end{align}
\end{corollary}
\textit{Proof:} See Appendix~\ref{App2}.

The above results demonstrate that the ISI term~$\mathbf{I}$ induces a structured, symbol-dependent randomness at the receiver side. 
In essence, the deterministic offset~$\tau$ introduces sufficient variability in the received amplitude distribution to contribute to privacy, even in the absence of externally added noise. 
The resulting DP properties of this natural perturbation will be analyzed in the following.

The crossover probabilities corresponding to bit detection errors in the presence of ISI are obtained as
\begin{align}\label{eq:crossover}
\zeta_1 &= \Pr(\hat{a}_1 = 1 \mid a_1 = -1)
= \mathds{E}_{p_{\mathbf{I}}}\!\left[\,Q\!\left(\frac{g(\tau) - \mathbf{I}}{\sigma}\right)\right],\\
\zeta_2 &= \Pr(\hat{a}_1 = -1 \mid a_1 = 1)
= \mathds{E}_{p_{\mathbf{I}}}\!\left[\,Q\!\left(\frac{g(\tau) + \mathbf{I}}{\sigma}\right)\right],
\end{align} 
where $\zeta_1$ and $\zeta_2$ represent the conditional error probabilities corresponding to transmitting $-1$ and $+1$, respectively, averaged over the random realizations of the ISI term~$\mathbf{I}$.

Given the asymmetric crossover probabilities $\zeta_1$ and $\zeta_2$, the effective privacy level can be computed from the definition of differential privacy as
\begin{align}\nonumber
\varepsilon 
&= \ln\!\Bigg(\max\!\Bigg\{
\frac{\zeta_1}{1 - \zeta_2},\,
\frac{1 - \zeta_2}{\zeta_1},\,
\frac{\zeta_2}{1 - \zeta_1},\,
\frac{1 - \zeta_1}{\zeta_2}
\Bigg\}\Bigg)\\[2pt]\label{privacy}
&= \ln\!\left(\frac{1 - \max\{\zeta_1, \zeta_2\}}{\min\{\zeta_1, \zeta_2\}}\right),
\end{align}
where the simplification in~\eqref{privacy} follows from the identity $Q(-x) = 1 - Q(x)$, which ensures that $\zeta_1 + \zeta_2 < 1$. 
The expression above provides a closed-form relationship between the crossover asymmetry and the achieved privacy budget~$\varepsilon$.

The next lemma provides analytical series representations for the expectations in~\eqref{eq:crossover}, obtained by expanding the $Q$-function and substituting the moment expressions of~$\mathbf{I}$ derived in Corollary~\ref{cor1}.

\begin{lemma}\label{lemma2}
The crossover probabilities in~\eqref{eq:crossover} can be expressed as
\begin{align}
\mathds{E}_{p_{\mathbf{I}}}\!\left[Q\!\left(\frac{g(\tau) - \mathbf{I}}{\sigma}\right)\right]
&= \frac{1}{2} - \frac{1}{\sqrt{2\pi}\,\sigma}
\sum_{n=0}^{\infty} \sum_{k=0}^{2n+1} \sum_{j=0}^{k}
\frac{(2n-1)!!\, g(\tau)^{2n+1-k} (-1)^{n+k}\, \mu_j\, \mu^{k-j}}
{\sigma^{2n} (2n+1-k)!\, j!\, (k-j)!},\\[4pt]
\mathds{E}_{p_{\mathbf{I}}}\!\left[Q\!\left(\frac{g(\tau) + \mathbf{I}}{\sigma}\right)\right]
&= \frac{1}{2} - \frac{1}{\sqrt{2\pi}\,\sigma}
\sum_{n=0}^{\infty} \sum_{k=0}^{2n+1} \sum_{j=0}^{k}
\frac{(2n-1)!!\, g(\tau)^{2n+1-k} (-1)^{n}\, \mu_j\, \mu^{k-j}}
{\sigma^{2n} (2n+1-k)!\, j!\, (k-j)!},
\end{align}
where $\mu$ and $\mu_j$ denote the raw and central moments of the ISI term~$\mathbf{I}$, respectively.
\end{lemma}

\textit{Proof:} See Appendix~\ref{App3}.

\section{Numerical Results}

In this section, we numerically illustrate the privacy–performance tradeoffs of the proposed mechanisms and validate the analytical developments of Sections III-A–III-C. In particular, we compare privacy obtained via deterministic phase rotation and ISI, and examine the role of the input distribution parameter $p$.

\subsection{Privacy via Phase Rotation}

\begin{figure}[ht]
 \centering 
 \scalebox{0.6} 
 {\includegraphics{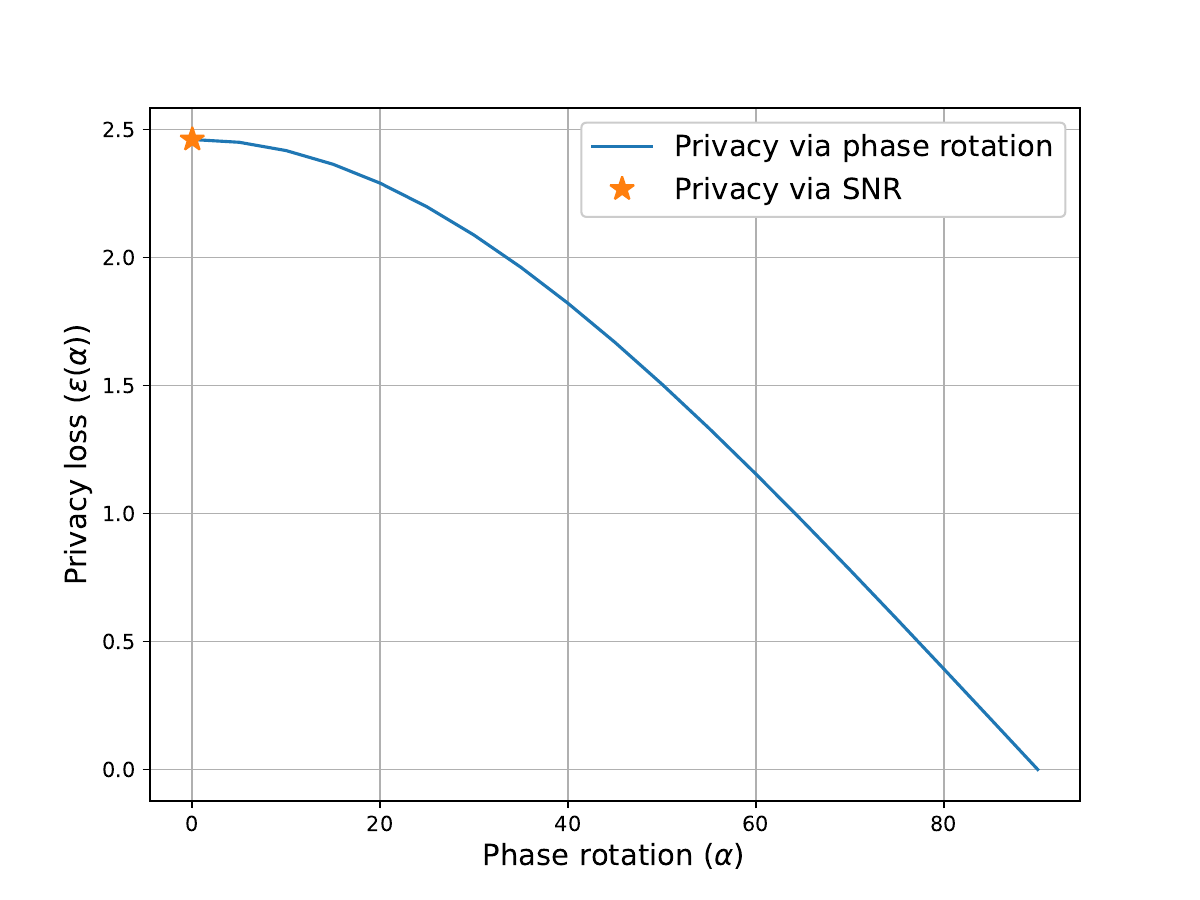}} 
 \caption{Privacy loss ($\varepsilon$) as a function of the rotation angle ($\alpha$), with $\alpha$ ranging from $0$ to $\pi/2$. }
 \label{Fig_fig1} 
\end{figure}

Figure~\ref{Fig_fig1} depicts the privacy loss $\varepsilon$ as a function of the phase rotation angle $\alpha$, where $\alpha$ varies from $0$ to $\pi/2$. The operating point at $\alpha = 0$, marked by a star, corresponds to the baseline privacy level achieved solely through the channel SNR, i.e., without any additional privacy-inducing mechanism beyond AWGN.

As $\alpha$ increases, the effective projection of the transmitted BPSK symbols onto the receiver’s detection axis is reduced by a factor of $\cos \alpha$, leading to a monotonic increase in the BER. Consequently, the privacy loss $\varepsilon$ decreases monotonically with $\alpha$, indicating progressively stronger privacy guarantees at the receiver. In the limiting case as $\alpha \to \pi/2$, the received signal becomes indistinguishable from noise along the detection axis, yielding $\varepsilon \approx 0$ and hence near-perfect privacy.

An important observation is that the curve in Figure~\ref{Fig_fig1} is independent of the input distribution parameter $p$. This follows directly from the fact that phase rotation affects the received signal geometry uniformly, regardless of the probability mass assigned to the transmitted symbols. Therefore, for any input distribution, the same privacy–rotation tradeoff applies.

\subsection{Privacy via ISI}

\begin{figure}[ht]
 \centering 
 \scalebox{0.6} 
 {\includegraphics{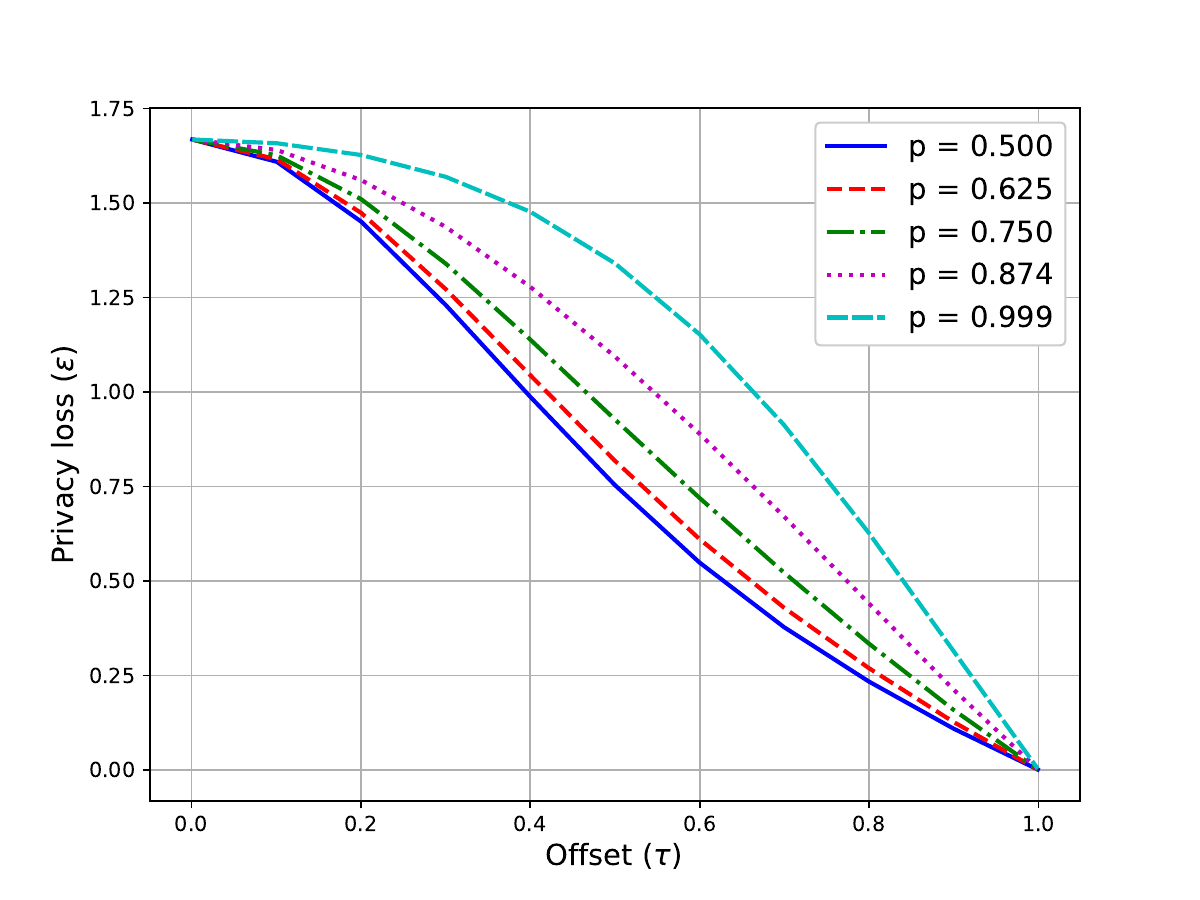}} 
 \caption{Privacy loss ($\varepsilon$) as a function of the ISI offset parameter ($\tau$), with $\tau \in [0,, T=1]$, for different values of the input distribution parameter $p$.}
 \label{Fig_fig2} 
\end{figure}

Figure~\ref{Fig_fig2} illustrates the privacy loss $\varepsilon$ as a function of the intentional timing offset $\tau \in [0, T=1]$, for different values of the input distribution parameter $p$. Introducing a nonzero $\tau$ creates deterministic ISI at the receiver, which manifests as a random perturbation due to the superposition of neighboring symbols weighted by the pulse-shaping function.

As $\tau$ increases, the ISI power grows, resulting in a higher BER at the receiver. Since DP is directly linked to the indistinguishability of the received outputs corresponding to neighboring input datasets, this increase in BER leads to a monotonic decrease in the privacy loss $\varepsilon$. At $\tau = 1$, the desired symbol component vanishes for sinc-based pulse shaping, and the received sample is dominated by ISI and noise. In this regime, the receiver is unable to infer the transmitted symbol, yielding $\varepsilon = 0$, which corresponds to perfect privacy.

The numerical results further reveal that, for any fixed value of $\tau$, the privacy loss is maximized when the input distribution is uniform, i.e., $p = 0.5$. This behavior is consistent with worst-case DP analysis, where the uniform distribution maximizes uncertainty and minimizes the asymmetry between conditional error probabilities. As $p$ deviates from $0.5$, the privacy loss decreases for the same ISI level, indicating stronger privacy guarantees for biased input distributions.

\subsection{Privacy–BER Tradeoff}

\begin{figure}[ht]
 \centering 
 \scalebox{0.6} 
 {\includegraphics{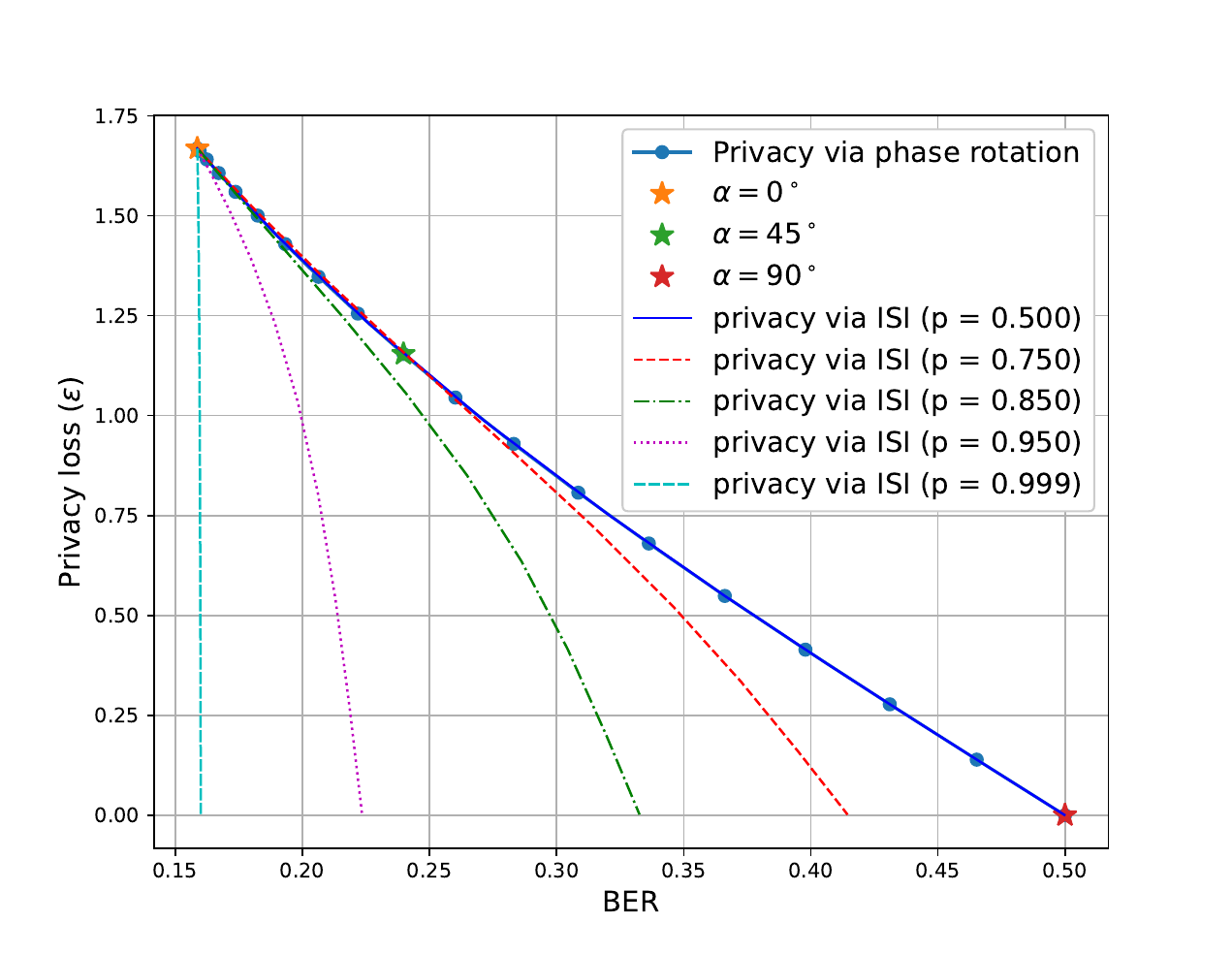}} 
 \caption{Privacy level ($\varepsilon$) as a function of the bit error rate (BER) for both phase rotation and intentional intersymbol interference.}
 \label{Fig_fig3} 
\end{figure}

Figure~\ref{Fig_fig3} presents the privacy loss $\varepsilon$ as a function of the BER for both privacy via phase rotation and privacy via intentional ISI, with multiple values of the input distribution parameter $p$. This figure provides a unifying comparison between the two privacy mechanisms in terms of their fundamental privacy–reliability tradeoff.

A particularly noteworthy numerical observation is that, for $p = 0.5$, the privacy–BER curve corresponding to intentional ISI coincides exactly with that of privacy via phase rotation. This is intuitively justified as follows. When $p = 0.5$, the equivalent channel between the input data and estimated ouput is a BSC($\zeta_1 = \zeta_2=\zeta(\tau)$). Therefore (\ref{privacy}) simplifies to $\ln{(1-\zeta(\tau))/\zeta(\tau)}$. For any given $\tau$, there is an $\alpha$ that achieves the same $\zeta(\tau)$ and hence same level of DP. This implies that, under a uniform input distribution, both mechanisms induce an equivalent stochastic mapping from transmitted bits to receiver decisions, resulting in identical privacy guarantees for a given BER.

For $p \neq 0.5$, however, intentional ISI exhibits a clear advantage. Specifically, for a given privacy loss $\varepsilon$, the ISI-based approach achieves the same privacy with a lower BER compared to phase rotation or SNR-based privacy. Equivalently, for a fixed BER, privacy via ISI yields a smaller $\varepsilon$, i.e., stronger privacy guarantees. This improvement becomes more pronounced as $p$ increases.

In the extreme case where $p \to 1$, the privacy–BER curve degenerates into a vertical line at $\mathrm{BER} \approx 0.15$, indicating that variations in ISI strength do not affect the BER while still enabling a wide range of achievable privacy levels. This regime highlights a key advantage of intentional ISI: for highly biased input distributions, significant privacy gains can be obtained without incurring additional reliability degradation.

Overall, Figure~\ref{Fig_fig3} demonstrates that privacy via intentional ISI not only generalizes existing SNR- and phase-based privacy mechanisms, but can also strictly outperform them for non-uniform input distributions by exploiting structured physical-layer interference rather than artificial noise.

\textit{Effect of skewed input distributions and limitations of DP}: It is well known that DP guarantees bound worst-case distinguishability between neighboring datasets, but do not preclude accurate inference when the prior distribution of the sensitive variable is highly skewed. In particular, when the input distribution is strongly biased (e.g., $p \approx 1$), an adversary can predict the dominant symbol with high confidence even in the absence of any observation, and DP protection does not fundamentally alter this fact. Consequently, under extreme bias, privacy metrics based on output indistinguishability may indicate weak protection, not due to a failure of the DP mechanism, but because the prior itself dominates the posterior inference. This effect is clearly observed in Figure \ref{Fig_fig3} for intentional ISI with $p=0.999$, where varying the timing offset $\tau$ drives the privacy parameter $\varepsilon$ toward zero. This analysis assumes an honest but curious receiver that applies the prescribed mismatched detector without attempting to estimate or compensate the unknown timing offset. A receiver equipped with timing recovery or sequence estimation capabilities could, in principle, infer $\tau$ from the received waveform and mitigate the induced ISI, thereby reducing the effective BER and weakening the achieved privacy. Accordingly, the privacy gains reported here should be interpreted within the assumed receiver model, and highlight the tradeoff between receiver knowledge, achievable utility, and privacy under highly nonuniform input distributions.

\section{Conclusion}
This paper investigated physical layer privacy mechanisms based on both modulation rotation and intentional ISI, with the aim of quantifying the resulting privacy-reliability tradeoffs in terms of an $\varepsilon$-based indistinguishability metric and the corresponding BER. By jointly analyzing these two approaches within a unified framework, we highlighted how deterministic waveform manipulations can be leveraged to control information leakage without relying on artificial noise injection.

Our results show that, unlike conventional SNR- and phase-based privacy mechanisms, intentional ISI can strictly outperform existing approaches under nonuniform input distributions. In particular, for highly biased sources, significant privacy gains can be achieved with little or no additional reliability degradation, enabling a wide range of achievable privacy levels at approximately fixed BER.

\appendix
\section{Proof of Lemma \ref{lemma1}}\label{App1}
If $\tau\in\{0,1\}$, the proof is obvious. Assume $\tau\in(0,1)$, and recall $T=1$. For $n=1$, we have
\begin{align}
\sum_{t\neq0}g(t+\tau)& = \frac{\sin\pi \tau}{\pi}\sum_{t\neq 0}\frac{(-1)^t}{t+\tau}\nonumber\\
      &=\frac{\sin\pi \tau}{\pi}\left(\sum_{t=1}^{+\infty}\frac{(-1)^t}{t+\tau}-\sum_{t=1}^{+\infty}\frac{(-1)^t}{t-\tau}\right)\label{note3}\\
      &=-\frac{\sin\pi \tau}{\pi}\left(\int_0^1\frac{x^{\tau}}{1+x}dx-\int_0^1\frac{x^{-\tau}}{1+x}dx\right)\label{id1}\\
      &=-\frac{\sin\pi \tau}{\pi}\left(\int_0^1\frac{x^{\tau}-x^{-\tau}}{1+x}dx\right)\nonumber\\
      &=-\frac{\sin\pi \tau}{\pi}(\frac{1}{\tau}-\frac{\pi}{\sin\pi \tau})\label{id2}\\
      &=1-\sinc(\tau),
    \end{align}
where the separation in (\ref{note3}) is permitted since bothe series converge, and (\ref{id1}) and (\ref{id2}) follow from the following two identities \cite{tableintegral}:
\begin{align}
    \sum_{t=1}^{+\infty}\frac{(-1)^{n+1}}{n+t}&=\int_0^1\frac{x^t}{1+t},\ t>-1\\
    \int_0^1\frac{x^{t}-x^{-t}}{1+x}dx &=\frac{1}{t}-\frac{\pi}{\sin\pi t},\ t\in(-1,1).
\end{align}
For $n>1$, we can write
\begin{align}
      \sum_{t\neq0}g(t-\tau)^n&=\sum_{t\neq0}g(t+\tau)^n\nonumber \\\label{app:0}
      &=\left(\frac{\sin{\pi \tau}}{\pi}\right)^n\sum_{t\neq 0}\left(\frac{(-1)^{t}}{t+\tau}\right)^n
\end{align}
We analyze the summation in (\ref{app:0}) for even and odd $n$, separately. For even values of $n\geq 2$ we have 
\begin{align}
\sum_{t\neq 0}\left(\frac{(-1)^{t}}{t+\tau}\right)^n &= \sum_{t=-\infty}^{0}\frac{1}{(t+\tau)^n}+\sum_{t=0}^{\infty}\frac{1}{(t+\tau)^n}-\frac{2}{\tau^n}\\
&= \sum_{t=0}^{\infty}\frac{1}{(t-\tau)^n}+\sum_{t=0}^{\infty}\frac{1}{(t+\tau)^n}-\frac{2}{\tau^n}\\
&= \zeta(n, -\tau)+\zeta(n, \tau)-\frac{2}{\tau^n}
\end{align}
For odd values of $n\geq 3$ we have 
\begin{align}
\sum_{t\neq 0}\left(\frac{(-1)^{t}}{t+\tau}\right)^n &= \sum_{t=-\infty}^{0}\frac{(-1)^{t}}{(t+\tau)^n}+\sum_{t=0}^{\infty}\frac{(-1)^{t}}{(t+\tau)^n}-\frac{2}{\tau^n}\\
&=-\sum_{t=0}^{\infty}\frac{(-1)^{t}}{(t-\tau)^n}+\sum_{t=0}^{\infty}\frac{(-1)^{t}}{(t+\tau)^n}-\frac{2}{\tau^n}\\
&=-\zeta^*(n, -\tau)+\zeta^*(n, \tau)-\frac{2}{\tau^n}
\end{align}
Therefore, for all $n$, we have 
\begin{align}
    \sum_{t\neq 0}g(t-\tau)^n &= \begin{cases} 
  1-\sinc(\tau) & \text{if } n =1 \\
  \left(\frac{\sin{\pi \tau}}{\pi}\right)^n(\zeta(n, \tau)+\zeta(n, -\tau)-\frac{2}{\tau^n}) & \text{if }  n \text{ is even}\\
  \left(\frac{\sin{\pi \tau}}{\pi}\right)^n(\zeta^*(n, \tau)-\zeta^*(n, -\tau)-\frac{2}{\tau^n}) & \text{if } n \text{ is odd}
\end{cases}
\end{align}

\section{Proof of Corollary \ref{cor1}}\label{App2}
Using the result (\ref{eq:lemma}) in Lemma \ref{lemma1}, we have
\begin{align}
\mu_n &= \sum_{t\neq 0}g(t-\tau)^n\mathds{E}[(\pmb{a}_t-(2p-1))^n]\\
&=G(n, \tau)2^n(p(1-p)^n+(1-p)(-p)^n)
\end{align} 
The higher moments $\mu'_n,\:(n\geq 2)$ around the origin are obtained as
\begin{align}
\mu'_n&= \mathds{E}[(\mathbf{I}-\mu+\mu)^n]\\
&=\sum_{j=0}^{n}\binom{n}{j} \mu_j \mu^{n-j}.
\end{align}

\section{Proof of Lemma \ref{lemma2}}\label{App3}

$Q(x)$ can be expanded using Taylor serious as below
\begin{align}
Q(x) &= \frac{1}{2} - \frac{1}{\sqrt{2\pi}} \int_{0}^{x} e^{-t^2/2} \, dt\\ 
&= \frac{1}{2} - \frac{1}{\sqrt{2\pi}} \int_{0}^{x} \sum_{n=0}^{\infty} \frac{1}{n!} \left(-\frac{t^2}{2}\right)^n \, dt\\
&= \frac{1}{2} - \frac{1}{\sqrt{2\pi}} \sum_{n=0}^{\infty} \frac{1}{n!} \left(-\frac{1}{2}\right)^n \int_{0}^{x} t^{2n} \, dt\\
&= \frac{1}{2} - \frac{1}{\sqrt{2\pi}} \sum_{n=0}^{\infty} \frac{1}{n!} \left(-\frac{1}{2}\right)^n \frac{x^{2n+1}}{2n+1}.
\end{align}
Therefore, for the expectation $\mathds{E}_{p_{\mathbf{I}}}\left[Q\left(\frac{g(\tau) - \mathbf{I}}{\sigma}\right)\right]$, we have 
\begin{align}
\mathds{E}_{p_{\mathbf{I}}}\left[Q\left(\frac{g(\tau) - \mathbf{I}}{\sigma}\right)\right]&=\mathds{E}_{p_{\mathbf{I}}}\left[\frac{1}{2} - \frac{1}{\sqrt{2\pi}\sigma} \sum_{n=0}^{\infty} \frac{1}{n!} \left(-\frac{1}{2}\right)^n \frac{(g(\tau)-\mathbf{I})^{2n+1}}{(2n+1)\sigma^{2n}}\right]\\
&=\mathds{E}_{p_{\mathbf{I}}}\left[\frac{1}{2} - \frac{1}{\sqrt{2\pi}\sigma} \sum_{n=0}^{\infty} \sum_{k=0}^{2n+1}\frac{1}{n!} \left(-\frac{1}{2}\right)^n\binom{2n+1}{k}
 \frac{g(\tau)^{2n+1-k}(-\mathbf{I})^{k}}{(2n+1)\sigma^{2n}}\right]\\
 &=\frac{1}{2} - \frac{1}{\sqrt{2\pi}\sigma} \sum_{n=0}^{\infty} \sum_{k=0}^{2n+1}\frac{1}{n!} \left(-\frac{1}{2\sigma^2}\right)^n\binom{2n+1}{k}
 \frac{g(\tau)^{2n+1-k}(-1)^k\mu'_k}{(2n+1)}\\
 &=\frac{1}{2} - \frac{1}{\sqrt{2\pi}\sigma} \sum_{n=0}^{\infty} \sum_{k=0}^{2n+1}\left(-\frac{1}{2\sigma^2}\right)^n\frac{(2n)!g(\tau)^{2n+1-k}(-1)^k\mu'_k}{n!k!(2n+1-k)!}\\
 &=\frac{1}{2} - \frac{1}{\sqrt{2\pi}\sigma} \sum_{n=0}^{\infty} \sum_{k=0}^{2n+1} \sum_{j=0}^{k}\left(-\frac{1}{2\sigma^2}\right)^n\frac{(2n)!g(\tau)^{2n+1-k}(-1)^k\mu_j\mu^{k-j}}{n!(2n+1-k)!j!(k-j)!}\\
 &=\frac{1}{2} - \frac{1}{\sqrt{2\pi}\sigma} \sum_{n=0}^{\infty} \sum_{k=0}^{2n+1} \sum_{j=0}^{k}\frac{(2n-1)!!g(\tau)^{2n+1-k}(-1)^{n+k}\mu_j\mu^{k-j}}{\sigma^{2n}(2n+1-k)!j!(k-j)!}
\end{align}
where we use the identity $\frac{(2n)!}{n!}=2^n n! (2n-1)!!$. Similarly for the other expectation $\mathds{E}_{p_{\mathbf{I}}}\left[Q\left(\frac{g(\tau) + \mathbf{I}}{\sigma}\right)\right]$, we have 
\begin{align}
\mathds{E}_{p_{\mathbf{I}}}\left[Q\left(\frac{g(\tau) + \mathbf{I}}{\sigma}\right)\right]&=\frac{1}{2} - \frac{1}{\sqrt{2\pi}\sigma} \sum_{n=0}^{\infty} \sum_{k=0}^{2n+1}\frac{1}{n!} \left(-\frac{1}{2\sigma^2}\right)^n\binom{2n+1}{k}
 \frac{g(\tau)^{2n+1-k}\mu'_k}{(2n+1)}\\
&=\frac{1}{2} - \frac{1}{\sqrt{2\pi}\sigma} \sum_{n=0}^{\infty} \sum_{k=0}^{2n+1}\left(-\frac{1}{2\sigma^2}\right)^n\frac{(2n)!g(\tau)^{2n+1-k}\mu'_k}{n!k!(2n+1-k)!}\\
&=\frac{1}{2} - \frac{1}{\sqrt{2\pi}\sigma} \sum_{n=0}^{\infty} \sum_{k=0}^{2n+1}\sum_{j=0}^{k}\frac{(2n-1)!!g(\tau)^{2n+1-k}(-1)^n\mu_j\mu^{k-j}}{\sigma^{2n}(2n+1-k)!j!(k-j)!}
\end{align}

\bibliography{REF}
\bibliographystyle{IEEEtran}
\end{document}